\documentstyle{article}
\title{
\bf Evidence for a first order transition\\
 in a plaquette 3d Ising-like action
}

\author{ {\it D. Espriu} \\
	 DECM and IFAE, Universitat de Barcelona\\
         Diagonal 647\\
         08028 Barcelona\\
         Spain\\
         \\
         {\it M. Baig}\\
	 IFAE, Universitat Aut\`onoma de Barcelona\\
	 Edifici C\\
	 08193 Bellaterra\\
	 Spain\\
	 \\
        {\it D.A. Johnston} \\
         and \\
	 {\it Ranasinghe P. K. C. Malmini $^{(a)}$}\\
         Dept. of Mathematics\\
         Heriot-Watt University\\
	 Riccarton,
         Edinburgh, EH14 4AS\\
         Scotland}
\begin{document}
  \maketitle
                      {\Large
                      \begin{abstract}
%
We investigate a 3d
Ising action
which corresponds to a
a class of
models defined by Savvidy and Wegner,
originally intended as discrete versions of string
theories on cubic
lattices. These models have vanishing bare surface tension and the
couplings are tuned in such a way that the action depends only
on the angles of the discrete surface, i.e. on the way the surface
is embedded in ${\bf Z}^3$. Hence the name gonihedric by which they
are known. We
show that the model displays 
a rather clear first order phase transition
in the limit where self-avoidance is
neglected and the action becomes a plaquette one.
This transition 
persists for small values of the self avoidance coupling, but it turns
to second order when this latter parameter is further increased.
These results exclude the use of this type of action as models
of gonihedric random surfaces, at least in the limit where self avoidance is
neglected.
\\ \\
\\ \\ \\ \\ \\ \\ \\ \\
$(a)$ {\it Permanent Address:} \\
Department of Mathematics\\
University of Sri Jayewardenepura\\
Gangodawila, Sri Lanka.
%
                        \end{abstract} }
%
  \thispagestyle{empty}
%
%
  \newpage
%
                  \pagenumbering{arabic}

In three dimensions the Ising model can be
thought of as describing volumes of negative spins in a sea of
positive ones, or viceversa.
It is well known that the familiar Ising action weights such configurations
according to the number of `broken' links, i.e. with an action which is
approximately proportional to the area of such volumes.
Apart from entropy effects, the energy of a given configuration depends
only on the surface, thus leading to a non-zero bare surface tension, but
it is independent of the `form' of the surface, i.e. of the precise way
the surface is embedded in the lattice. A rough surface weighs exactly the
same as a smooth one as long as the area is the same.

Is it possible to
define a Ising-like action where precisely the
opposite situation takes place? That is, is it possible to assign weights to
a Ising action in such a way that smooth surfaces are preferred? The extreme
case of such a situation would be one where the bare surface tension would
be zero and, if,
at all, a renormalized surface tension would be generated by fluctuations.

Savvidy et al. \cite{1}
recently answered the above question in the positive. They started
by suggesting a
a novel discretized random surface theory, the so-called gonihedric string,
whose action is
\begin{equation}
S = {1 \over 2} \sum_{<ij>} | \vec X_i - \vec X_j | \theta (\alpha_{ij}),
\label{e4a}
\end{equation}
where the sum is over the edges of some triangulated surface,
$\theta(\alpha_{ij}) = | \pi - \alpha_{ij} |^{\zeta}$,
$\zeta$ is some exponent,
and $\alpha_{ij}$ is the dihedral angle between 
neighbouring triangles with common link $<ij>$. 
This definition of the action was
inspired by the geometrical notion
the linear size
of a surface, originally defined by Steiner.
Some difficulties with the model for $\zeta=1$ were already
pointed out in \cite{jon}. Possible ways to cure these difficulties
are to add additional Gaussian terms\cite{esp} or simply to choose
$\zeta<1$,
for which the model is believed to be satisfactory, but perhaps lacking
a direct geometrical interpretation. Lattice models of random surfaces are
prone to this type of difficulties and one may wonder to what extent these
are caused by the discretization. Notice that
in eq. (\ref{e4a}) it is the surface itself that is
discretized, rather than the space in which it is embedded.

The next step
consists in
discretizing the embedding space itself by
restricting the allowed surfaces to be the plaquettes of a cubic
lattice (if we stay in three dimensions, as we will hereafter).
Savvidy and Wegner
\cite{7,8,8a,8b} rewrote the resulting theory
as a generalized Ising model by using
the geometrical spin cluster boundaries
to define the surfaces. This will be the model object of our interest here.

Savvidy and Wegner take the energy of a surface on a cubic
lattice to be given
by
$E=n_2 + 4 \kappa n_4$, where $n_2$ is the number
of links where two plaquettes meet at a right angle and
$n_4$ is the number of links where four plaquettes
meet at right angles. Thus flat surfaces are energetically favoured.
$\kappa$ is a free parameter which determines the relative
weight of a self-intersection of the surface.
In the limit $\kappa \rightarrow \infty$
the surfaces would be strongly self-avoiding,
whereas the opposite limit $\kappa \rightarrow 0$
would be that of phantom surfaces that could
pass through themselves without hindrance.
Naively this model would seem to correspond precisely to $\zeta=1$.
The contribution of the different configurations in this  model should be
contrasted with the familiar 3d Ising model with nearest neighbour
interactions where the surfaces contribute to the partition function
according to their areas. Here we have no area term at all.

On a cubic lattice 
the generalized gonihedric Ising Hamiltonian which
reproduces the desired energy 
$E=n_2 + 4 \kappa n_4$ contains nearest neighbour ($<i,j>$),
next to nearest neighbour ($<<i,j>>$) and round a plaquette ($[i,j,k,l]$)
terms
\begin{equation}
H= 2 \kappa \sum_{<ij>}^{ }\sigma_{i} \sigma_{j} -
\frac{\kappa}{2}\sum_{<<i,j>>}^{ }\sigma_{i} \sigma_{j}+
\frac{1-\kappa}{2}\sum_{[i,j,k,l]}^{ }\sigma_{i} \sigma_{j}\sigma_{k} \sigma_{l}.
\label{e1}  
\end{equation}
Such generalized Ising actions
have quite complicated phase structures
for generic choices of the couplings \cite{9,9a,10}. The particular
ratio of couplings in eq.(\ref{e1}), however, is non-generic. In fact
actions such as eq.(\ref{e1}) have a symmetry which is not present
for generic couplings
- it is possible to flip any plane
of spins at zero energy cost.
The phase diagram displays only a single transition
to a low temperature layered state that, as a consequence
of the flip symmetry, is equivalent to a ferromagnetic
ground state. Simulations of the model at $\kappa=1$
\cite{8,11} suggest, rather remarkably given that the
model is defined in {\it three dimensions}, that the critical exponents
and even the critical temperature are close to
the Onsager values of the {\it two-dimensional}
Ising model with nearest neighbour interactions.
In addition, the simulations of \cite{11}
indicated that this picture still held for
$\kappa>1$.

The nature of the transition for $\kappa < 1$ was rather less clear.
A zero temperature analysis \cite{11} shows that there is an
increased symmetry in the ground state when 
$\kappa=0$, which is already apparent from the Hamiltonian itself.
In effect, for $\kappa=0$ it is now
possible to flip diagonal spin planes
as well as those that are perpendicular to the lattice axes.

Although not a local gauge
symmetry, the flip symmetry of the model is intermediate between this
and a global symmetry.
This symmetry poses something of a problem when carrying out
simulations, as it means that a
simple ferromagnetic order parameter
\begin{equation}
M = \left< {1 \over L^3} \sum_i \sigma_i \right>.
\label{ord}
\end{equation}
will be zero in general, because of the layered
nature of the ground state. Even staggered
magnetizations would not do as the interlayer
spacing can be arbitrary.
On a finite lattice it is possible, however, to
force the model into the ferromagnetic
ground state, which is equivalent
to any of the layered ground states, with a suitable choice of
boundary conditions.

We have used a variety of boundary conditions to do the job.
We have, for instance, used fixed boundary conditions which
penalize
any flipped spin planes by a boundary term. We have also
considered mixed boundary conditions in which half the boundary of
the box is set to a prescribed value and the other half to the
opposite one. (This technique allows us to measure the renormalized
surface tension, but these results merit a separate analysis\cite{tension}.)
Another possibility
is to fix internal planes of spins
in the lattice, whilst retaining the periodic boundary conditions.
This has the desired effect of picking out the ferromagnetic
ground state, whilst minimizing any finite size effects.
With either fixed spin planes or boundaries we can therefore still employ
the simple order parameter of eq.(\ref{ord}).

For $\kappa=0$ the
Hamiltonian
we simulate is thus
\begin{equation}
H= \frac{1}{2}\sum_{[i,j,k,l]}^{ }\sigma_{i} \sigma_{j}\sigma_{k} \sigma_{l}.
\label{e2}
\end{equation}
In a sense, this Hamiltonian represents the opposite limit to the $\kappa=1$
case, where no plaquette term is present. In this case the
nearest neighbour and next to nearest neighbour terms are absent.
We emphasize that the spins live on the {\it vertices}
of the cubic lattice rather than the links, so the model of eq.(\ref{e2})
is not the three dimensional $Z_2$ gauge model that is dual to 
the three dimensional Ising model.

We shall now
describe without further ado our simulations and the results for $\kappa=0$
and then discuss to what extent these survive when we go to $\kappa\neq 0$.
For $\kappa=0$ we carried
out the bulk of the simulations on lattices of size $L=10^3,12^3,15^3,18^3$ and $20^3$.
as well as some further simulations for 
$L=22^3,25^3$. Periodic boundary conditions were imposed
in the three directions and three internal perpendicular planes
of spins fixed to be $+ 1$.
The update algorithm used was a simple Metropolis update and 
we carried out $4\times 10^4$ thermalization sweeps followed by
$10^5$
measurement sweeps at each $\beta$ value simulated.
We measured the usual thermodynamic quantities for the model:
the energy $E$, specific heat $C$,
(standard) magnetization $M$, susceptibility $\chi$
and various cumulants.
The zero temperature analysis shows that the large $\beta$ limit
of the energy is $-3/2$ which gives a useful check on the veracity
of the results. A glance at Fig.1 shows 
that this limit is approached satisfactorily and
that there is a strong
signal for a discontinuity in the energy,
and hence a first order transition, with increasing
lattice size at $\beta_c \simeq 0.505$. The emergence
of a discontinuity in the energy is reflected
in the measurements of the specific heat which
scales linearly with the volume. 

Relying directly on the energy,
or even properties like hysteresis, to divine the order of a
transition is a rather
perilous business but one usually reliable
indicator is the scaling 
of Binder's energy cumulant
\begin{equation}
U_E = 1 - {\langle E^4\rangle  \over 3 \langle E^2\rangle^2 }
\label{bind}
\end{equation}
which scales to 2/3 at a continuous transition
and a non-trivial value ($< 2/3$) at a 
first order transition.
In Fig.2 we show the minimum values of $U_E$
(which occur very close to the observed transition point)
at $\kappa=0$
for various lattice sizes. Although the error
bars for the various lattice sizes are large
due to both fluctuations 
close to the transition points and uncertainty
in determining the real minimum values, there are clear
indications of a non-trivial scaling (a second order
transition would be expected to approach the 
horizontal $U_E=2/3$ line smoothly with increasing lattice size).
We also plot the values of $U_E$ for $\kappa=0.1,0.5$ 
on smaller lattices, but with better statistics, discussed
below.

It is not only the energy which displays a discontinuity at $\beta \simeq 0.505$.
There is a very sharp change in the magnetization in this region too, 
as witnessed
by Fig.3. This is another generic
feature of first order transitions and
is also reflected in the very large peak that appears
in the susceptibility $\chi$ 
with increasing lattice size which also displays
trivial first order scaling.

Let us now turn to $\kappa\ne 0$.
The Hamiltonian of eq.(\ref{e1}) was shown in \cite{11} to display a
second order transition for $\kappa \ge 1$, so it is of some
interest to determine the nature of the crossover to
the first order behaviour. Our simulations at $\kappa=0.1$
were carried out on
lattices up to $18^3$,
using both fixed and mixed boundary conditions
in order to facilitate string tension measurements.
They show that $\beta_c$ is greater than the $\kappa \ge 1$ value
of $0.44$, but smaller than the one of $\kappa=0$, and that the
discontinuity in the energy is already
quite marked.
The minimum values of $U_E$ (Fig.2) provide
evidence for a first order transition at
$\kappa=0.1$. The difference with the asymptotic value expected for a
second order transition (2/3) is now much smaller than for
$\kappa=0$, but the smaller error bars due to the increased statistics
($10^6$ sweeps per $\beta$-value) make the evidence
for scaling to a non-trivial value of equivalent statistical significance.
Furthermore
it is clear from Fig.4 that the peak in the specific heat
is scaling linearly with the volume for both sets of boundary conditions,
as would be expected at a first order transition. As for $\kappa=0.5$
earlier simulations reported
briefly in \cite{11} suggested a second order transition
at $\beta_c=0.44$, albeit with a slightly larger specific heat peak
than at $\kappa=1$. We have now performed further
simulations at $\kappa=0.5$ confirming that
$U_E$ clearly scales to the second order value of 2/3. This is
also shown in Fig.2. A
non-trivial exponent for the specific heat scaling is also found.
It would thus appear that the gonihedric Ising model has a
sharp crossover to first order behaviour somewhere between 0.1 and 0.5.
\footnote{
An application of the cluster variational method to the model
arrives at similar conclusions \cite{gon}.}.

We can get some hints as to why 
there is a first order transition in the $\kappa=0$
model by considering the dual Hamiltonian \cite{7}.
This can be written in various forms, but the most
illuminating for our purposes is
\begin{equation}
H_{dual} = \sum_{\xi} \Lambda^{\chi} (\xi) \Lambda^{\chi} (\xi + \chi)
+ \Lambda^{\eta} (\xi) \Lambda^{\eta} (\xi + \eta)
+ \Lambda^{\zeta} (\xi) \Lambda^{\zeta} (\xi + \zeta)
\label{dual1}
\end{equation}
where $\Lambda^{\chi} = ( 1, 1, -1, -1)$, $\Lambda^{\eta} = ( 1, -1, 1, -1)$
and $\Lambda^{\zeta} = ( 1, -1, -1, 1)$ are one dimensional
irreducible representations of the fourth order Abelian group
and $\xi, \eta, \zeta$ are unit vectors in the dual lattice.
The spins may also be considered as Ising ($\pm 1$) spins
if we set $\Lambda^{\zeta} = \Lambda^{\chi} \; \Lambda^{\eta}$,
which gives the following Hamiltonian
\begin{equation} 
H_{dual} = \sum_{\xi} \Lambda^{\chi} (\xi) \Lambda^{\chi} (\xi + \chi)
+ \Lambda^{\eta} (\xi) \Lambda^{\eta} (\xi + \eta) 
+ \Lambda^{\chi} (\xi) \Lambda^{\eta} (\xi) \Lambda^{\chi} (\xi + \zeta)
\Lambda^{\eta} (\xi + \zeta).
\label{dual2}
\end{equation}
This is recognizable as an anisotropically coupled Ashkin-Teller model.
In particular, because all the couplings in front of the different terms
are equal, it looks like the four-state Potts model limit of
the Ashkin-Teller family. As isotropic $Q>2$ state Potts models
have a first order phase transition in 3d, it is perhaps not so
surprising that the particular anisotropic variant in
equ.(\ref{dual2}), and hence its dual in eq.(\ref{e2}) which we have simulated,
show a first order transition.
The dual Hamiltonian is only valid at $\kappa=0$ so 
there is no contradiction in the
continuous transition that has been observed 
in simulations at larger non-zero values
of $\kappa$.

To summarize, we have simulated the
gonihedric 3d Ising model suggested by
Savvidy and Wegner in the context of string theory
at an exceptional value of the self-avoidance coupling,
$\kappa=0$, where the theory possesses an enhanced symmetry.
We have found a first order transition and given some
arguments from the dual theory as to why this might be expected.
We have also observed in further simulations
at small $\kappa$ values that the
crossover to a continuous transition
is quite sharp.
The presence of some degree of self-avoidance, in the form
of a non-zero value for $\kappa$, would thus appear
to have an important influence on the universality
properties of the Savvidy-Wegner/gonihedric models.
Although the model at $\kappa=0$ is not a candidate
for constructing a continuum string theory
in three dimensions because it does
not possess a continuous transition, it still
constitutes an interesting addition to the bestiary
of Ising-like models.
It is possible to construct Savvidy-Wegner 
models in four and higher dimensions, so 
extending the investigations of the current paper
to four dimensions, which is more directly
relevant for (Euclideanized) relativistic string theory,
may be profitable.

\bigskip\bigskip

R.P.K.C. Malmini was supported by Commonwealth Scholarship 
SR0014. DE and DAJ were partially supported by EC HCM network
grant ERB-CHRX-CT930343. DE acknowledges the financial support of CICYT and
CIRIT through grants AEN95-0590 and GRQ93-1047,
respectively.

\vfill
\eject

\clearpage \newpage
\begin{figure}[htb]
\vskip 20.0truecm
\includegraphics{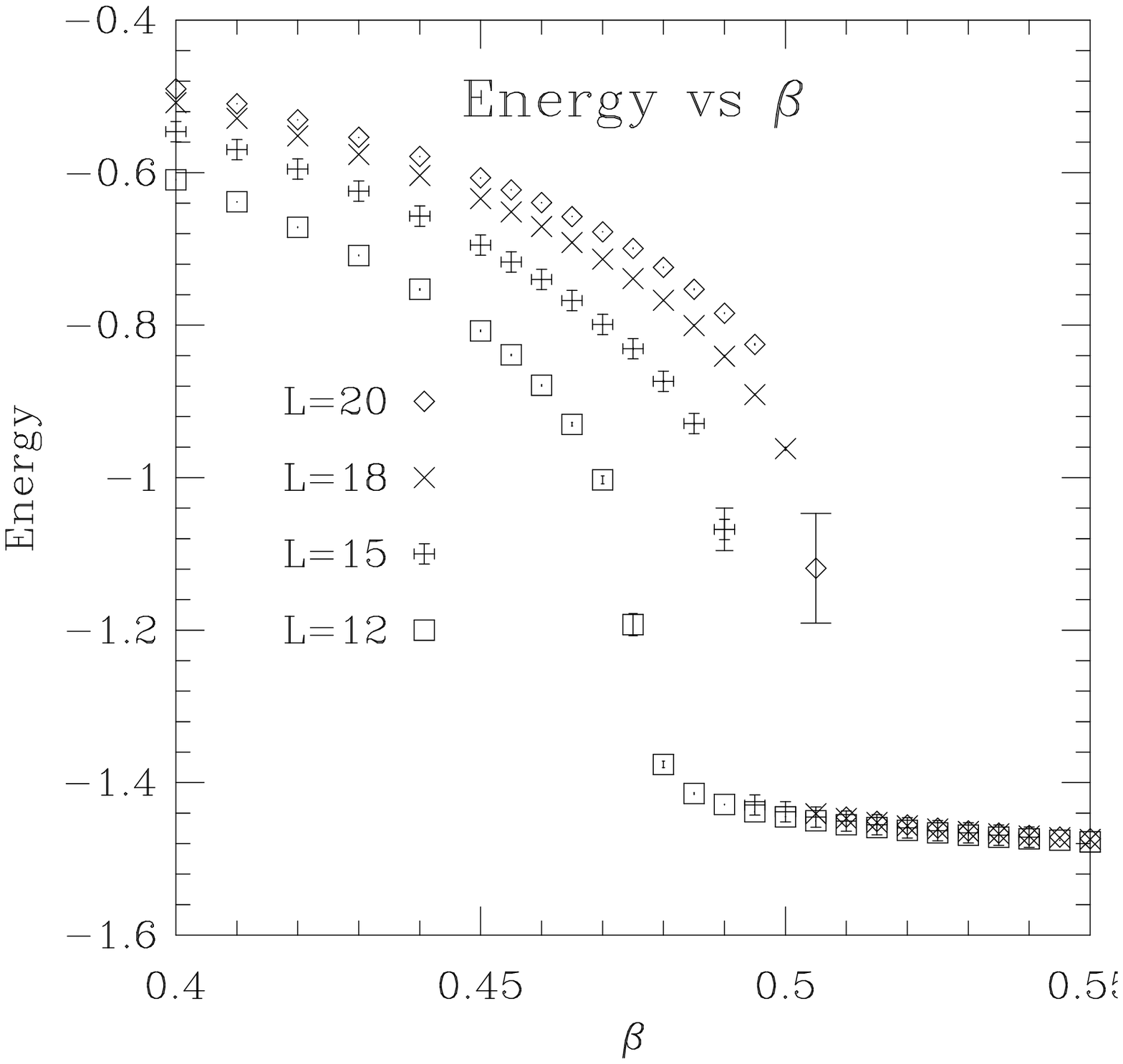}
\caption[2]{\label{fig1}The energy for various lattice sizes at $\kappa=0$}
\end{figure}
\clearpage \newpage
\begin{figure}[htb]
\vskip 20.0truecm
\includegraphics{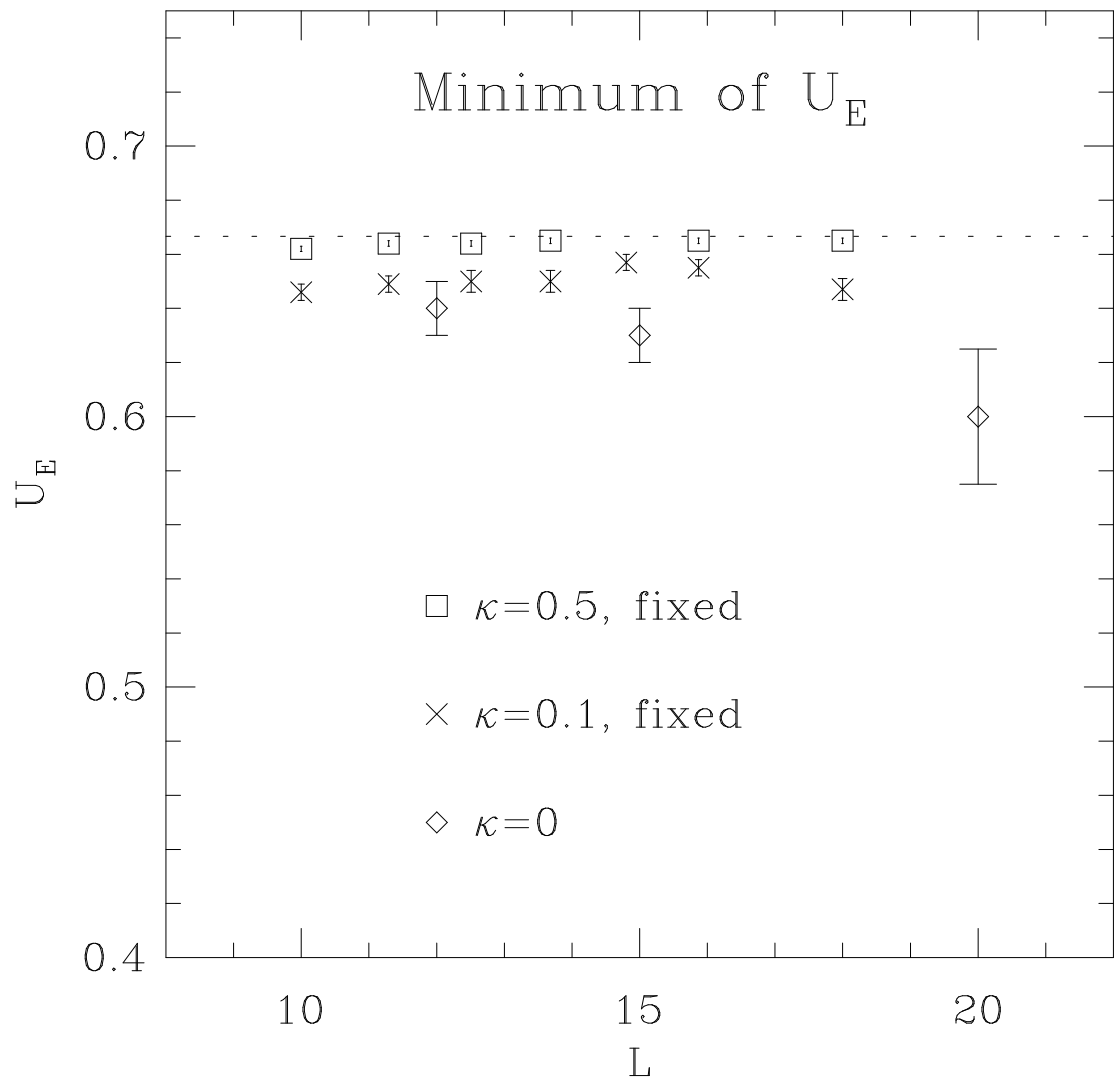}
\caption[2]{\label{fig2}The minimum value of $U_E$ for various lattice sizes.
The expected value of 2/3 at a second order transition is indicated as a 
horizontal line.}
\end{figure}
\clearpage \newpage
\begin{figure}[htb]
\vskip 20.0truecm
\includegraphics{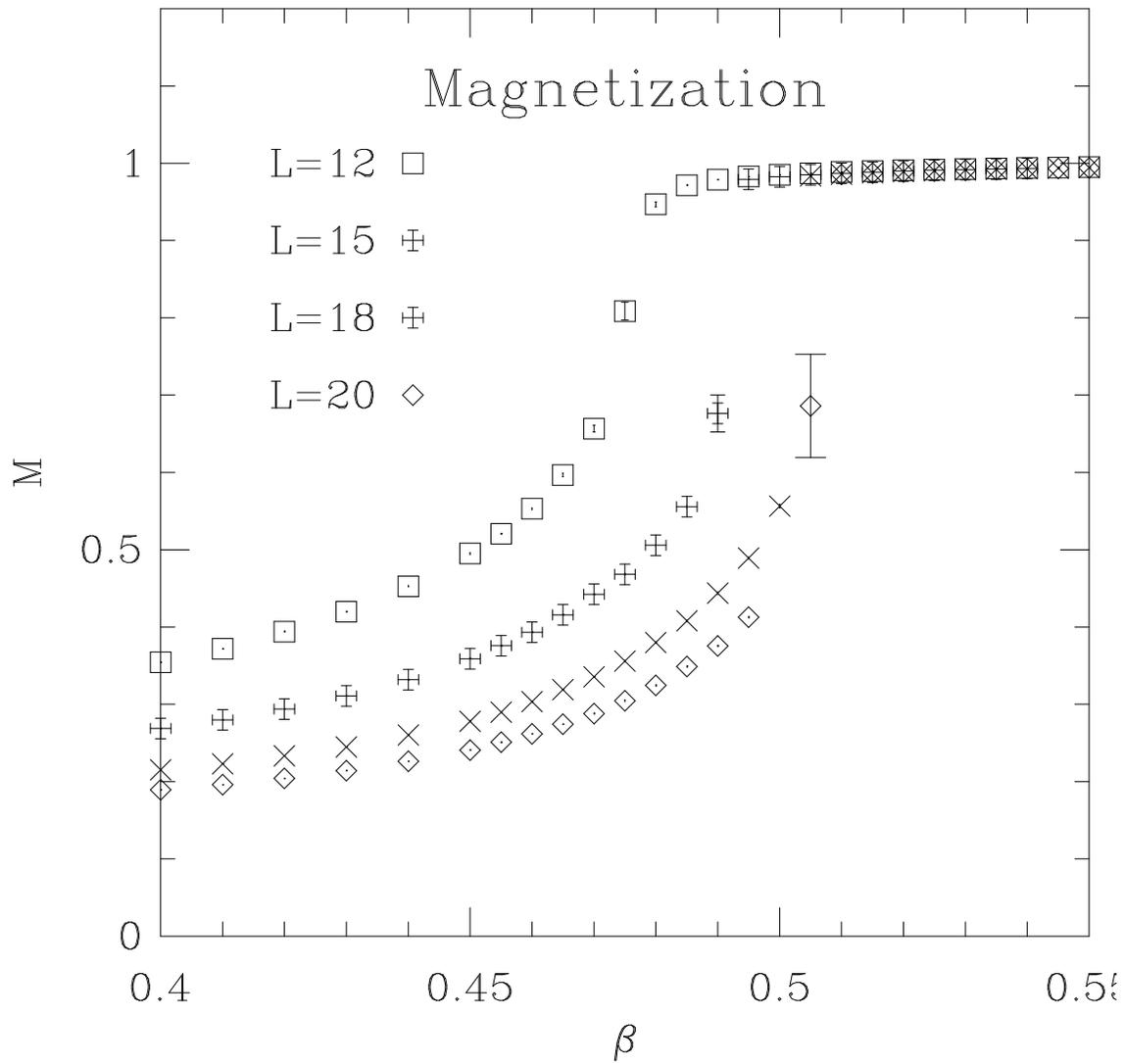}
\caption[2]{\label{fig3}The magnetization for various lattice sizes at
$\kappa=0$} \end{figure}
\clearpage \newpage
\begin{figure}[htb]
\vskip 20.0truecm
\includegraphics{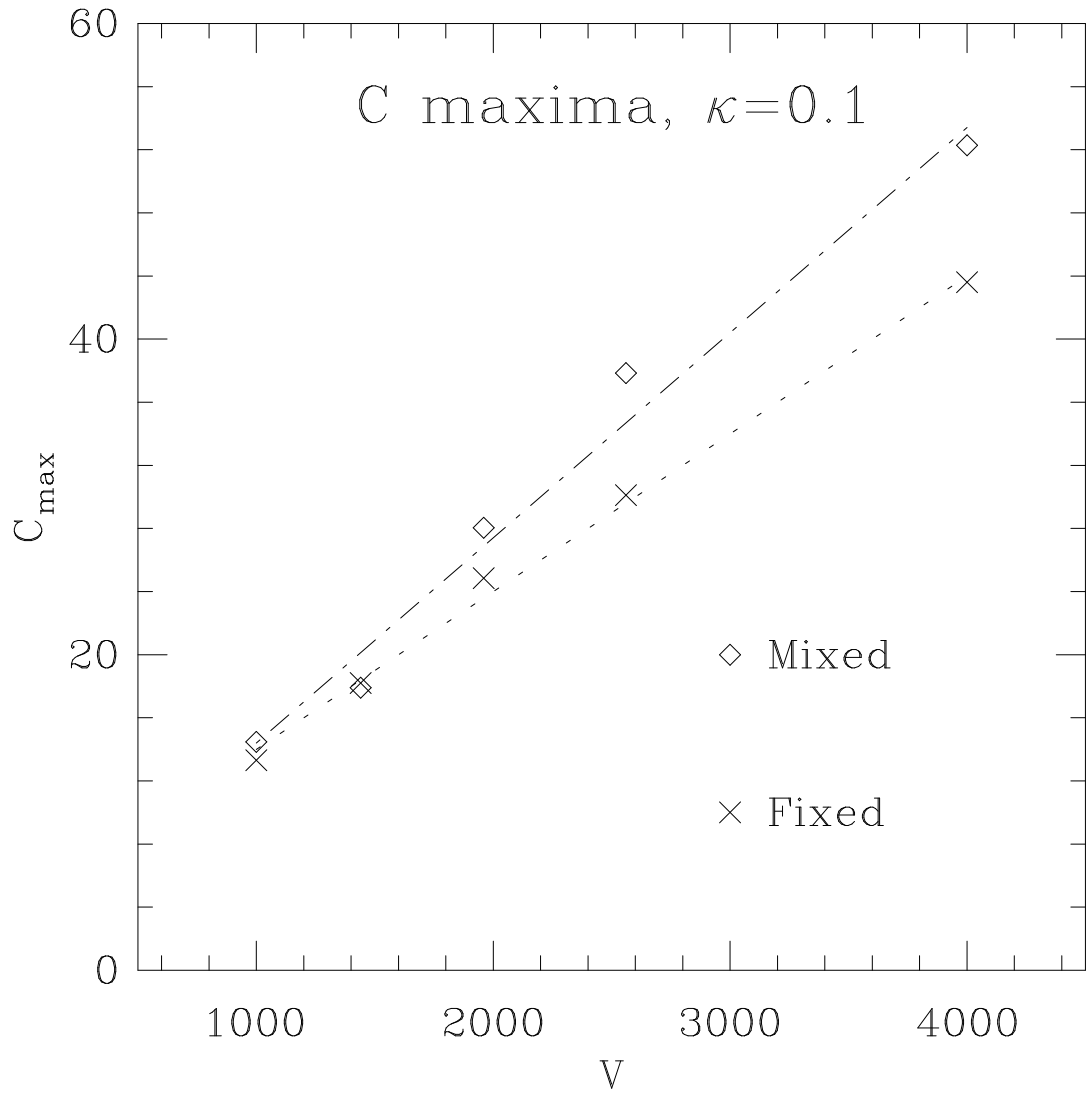}
\caption[2]{\label{fig4}The scaling of the specific heat maximum 
for the two sorts of boundary conditions used at $\kappa=0.1$. The best
fit lines of the form $A + B V$ are shown.}
\end{figure}

\end{document}